\pdfoutput=1

\documentclass[11pt]{article}

\usepackage{ACL2023}

\usepackage{times}
\usepackage{latexsym}

\usepackage[T1]{fontenc}

\usepackage[utf8]{inputenc}

\usepackage{microtype}

\usepackage{inconsolata}

\usepackage{booktabs}
\usepackage{multirow}
\usepackage{amsfonts} 
\usepackage{graphicx}
\usepackage{amsmath}
\usepackage{mathtools,amssymb,mathrsfs}
\usepackage{microtype}
\usepackage{enumitem}

\title{DopplerBAS: Binaural Audio Synthesis Addressing Doppler Effect}

\author{
Jinglin Liu\thanks{\quad Equal contribution.} 
$^1$ , 
Zhenhui Ye\footnotemark[1]
$^1$, 
Qian Chen$^2$,  
Siqi Zheng$^2$, 
Wen Wang$^2$, 
Qinglin Zhang$^2$
\\
\textbf{Zhou Zhao$^1$} \\
$^1$Zhejiang University \\
$^2$Speech Lab of DAMO Academy, Alibaba Group \\
}

\begin{document}
\maketitle

\begin{abstract}
Recently, binaural audio synthesis (BAS) has emerged as a promising research field for its applications in augmented and virtual realities. Binaural audio helps 
users orient 
themselves and establish immersion by providing the brain with interaural time differences reflecting spatial information. However, existing BAS methods are limited in terms of phase estimation, which is crucial for spatial hearing. In this paper, we propose the \textbf{DopplerBAS} method to explicitly address the Doppler effect of the moving sound source. Specifically, we calculate the radial relative velocity of the moving speaker in spherical coordinates, which further guides the synthesis of binaural audio. This simple method 
introduces no additional hyper-parameters and does not modify the loss functions, and is plug-and-play: it scales well to different types of backbones. DopperBAS distinctly improves the representative WarpNet and BinauralGrad backbones in the phase error metric and reaches a new state of the art (SOTA): 0.780 (versus the current SOTA 0.807). Experiments and ablation studies demonstrate the effectiveness of our method.
\end{abstract}

\section{Introduction}
Binaural audio synthesis (BAS), which aims to render binaural audio from the monaural counterpart, has become a prominent technology in artificial spaces (e.g. augmented and virtual reality)~\cite{Richard2021NeuralSO,Richard2022DeepIR,leng2022binauralgrad,Lee2022NeuralFS,Parida2022BeyondMT,Zhu2022BinauralRO,Park2022AMT}. Binaural rendering provides users with an immersive spatial and social presence~\cite{Hendrix1996TheSO,gao2019visualsound,Huang2022EndtoEndBS,Zheng2022InterpretableBR}, by producing stereophonic sounds with accurate spatial information. Unlike traditional single channel audio synthesis~\cite{OordDZSVGKSK16,ChenZZWNC21}, BAS places more emphasis on accuracy over sound quality, since humans need to interpret accurate spatial clues to locate objects and sense their movements consistent with visual input~\cite{Richard2021NeuralSO,Lee2022GlobalHI}. 

Currently, there are three types of neural networks (NN) to synthesize binaural audio. Firstly, \citet{Richard2021NeuralSO} collects a paired monaural-binaural speech dataset and provides an end-to-end baseline with geometric and neural warping technologies. Secondly, to simplify the task, \citet{leng2022binauralgrad} decompose the synthesis into a two-stage paradigm: the common information of the binaural audio is generated in the first stage, based on which the binaural audio is generated in the second stage.
They also propose to use the generative model DDPM~\cite{JonathanHo2020DenoisingDP} to improve the audio naturalness. Thirdly, to increase the generalization capability for the out-of-distribution audio, \citet{Lee2022NeuralFS} renders the speech in the Fourier space. These non-linear NN-based methods outperform the traditional digital signal processing systems based on a linear time-invariant system~\cite{Savioja1999CreatingIV,Zotkin2004RenderingLS,Sunder2015NaturalSR}.

However, these NN methods still have room for improvement in accuracy, especially phase accuracy. \citet{Richard2022DeepIR} claims that the correct phase estimation is crucial for binaural rendering~\footnote{Our ears can discriminate interaural time differences as short as 10$\mu$s~\cite{CPBrown1998ASM,Richard2021NeuralSO,johansson2022interaural}.}. Actually, the previous works tend to view the scene ``statically'', and only take into account the series of positions and head orientations. This motivates us to propose \textbf{DopplerBAS}, which facilitates phase estimation by explicitly introducing the Doppler effect~\cite{Gill1965TheDE,giordano2009college} into neural networks. Specifically, 1) we calculate the 3D velocity vector of the moving sound source in the Cartesian coordinates and then decompose this 3D velocity vector into a velocity vector in the spherical coordinates relative to the listener; 2) According to the Doppler effect, we use the radial relative velocity as an additional condition of the neural network, to incentivize the model to sense
the moving objects. We also analyze the efficacy of
different types of velocity conditions through extensive experiments. 

Naturally, DopplerBAS can be applied to different neural binaural renderers without tuning hyper-parameters. We pick two typical recent backbones to demonstrate the effectiveness of our method: 1) WarpNet~\cite{Richard2021NeuralSO}, a traditional neural network optimized by reconstruction losses; 2) BinauralGrad~\cite{leng2022binauralgrad}, a novel diffusion model optimized by maximizing the evidence bound of the data likelihood. Experiments on WarpNet and BinauralGrad are representative and could show the generalizability of our proposed DopplerBAS on other conditions based on gains on these two models. The contributions of this work can be summarized as follows:
\begin{itemize}
    \item We propose DopplerBAS, which distinctly improves WarpNet and BinauralGrad in the phase error metric and produces a new state of the art performance: 0.780 (vs. the current state of the art 0.807).
    \item We conduct analytical experiments under various velocity conditions and discover that: 1) NN does not explicitly learn the derivative of position to time (velocity); 2) The velocity condition is beneficial to binaural audio synthesis, even the absolute velocity in the Cartesian coordinates; 3) The radial relative velocity is the practical velocity component, which obeys the theory of the Doppler effect.
\end{itemize}





\section{Method}
In this work, we focus on the most basic BAS scenario where only the monaural audio, the series of positions and head orientations are provided~\cite{Richard2022DeepIR,leng2022binauralgrad}, rather than other scenarios where extra modalities~\cite{Xu2021VisuallyIB} are present. Note that scenarios with extra modalities present are different tasks. Also, as demonstrated in this paper, our proposed DopplerBAS is plug-and-play and can be easily integrated into other more complex scenarios. In this section, we will introduce the Doppler Effect as the preliminary knowledge, and then introduce the proposed method DopplerBAS. We will describe how to calculate and decompose the velocity vector, and how to apply this vector to two different backbones.

\subsection{Doppler Effect}
The Doppler effect~\cite{Gill1965TheDE} is the change in frequency of a wave to an observer, when the wave source is moving relative to it. This effect is originally used in radar systems to reveal 
the characteristics of interest for the target moving objects
~\cite{Chen2006MicroDopplerEI}. It can be formulated as:
\begin{equation}
\label{eq:doppler}
    f = \left( \frac{c}{c\pm v_{r}} \right) f_{0},
\end{equation}
where $c$, $v_{r}$, $f_0$ and $f$ are the propagation speed of waves, the radial relative velocity of the moving sound source, the original frequency of waves and the received frequency of waves, respectively.
 
\begin{figure}[!h]
	\centering
	\includegraphics[width=0.48\textwidth]{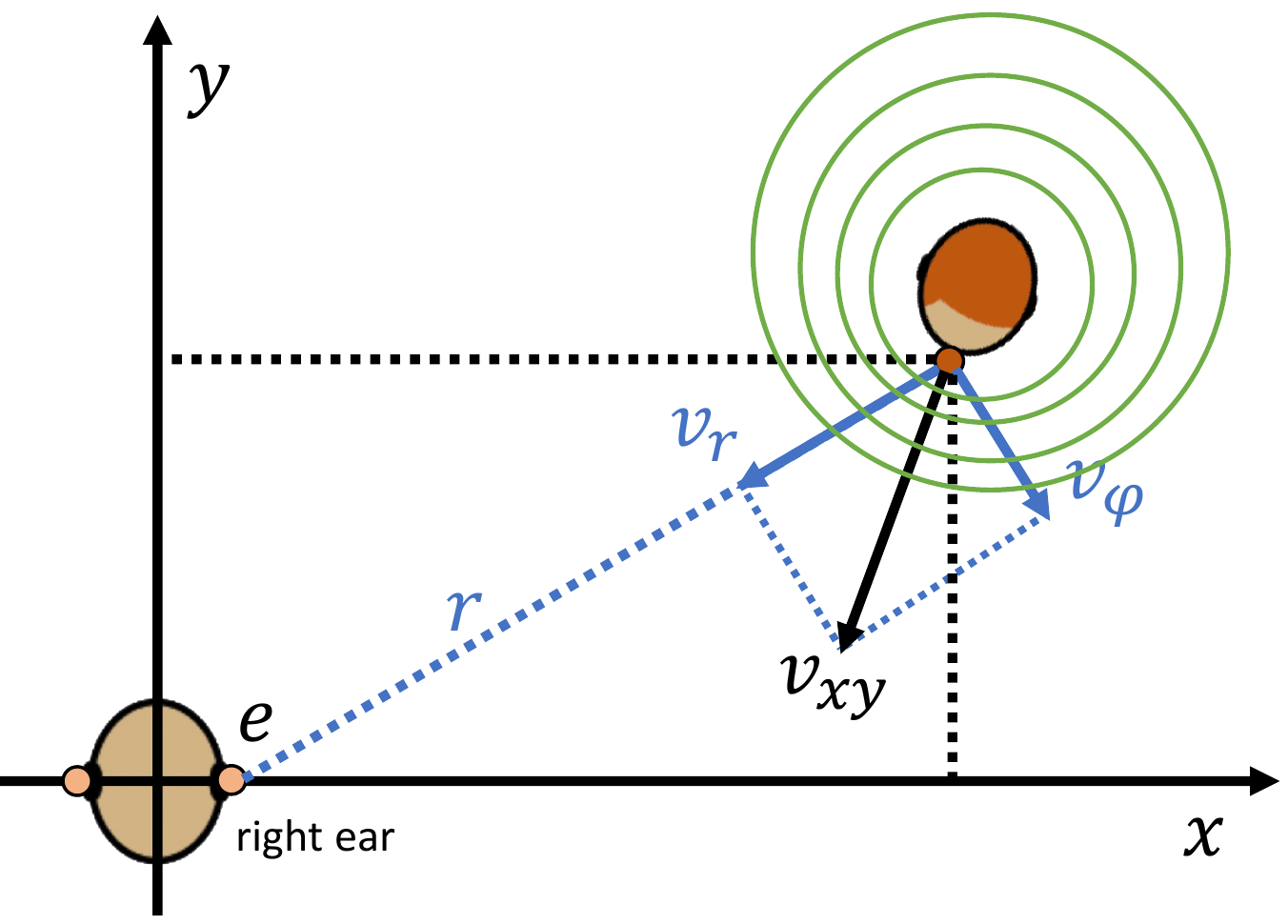}
	\caption{We illustrate the top view where the height dimension is omitted for simplicity. The sound source is moving in the x-y plane with the velocity $v_{xy}$. This velocity is decomposed into the radial velocity $v_r$ relative to one ear (e.g., the right ear).}
	\label{fig:velocity_decompose}
 \vspace{-2mm}
\end{figure}

\begin{table*}[!htb]
\small
\begin{center}
\begin{tabular}{l|cccccc}
\toprule
Model  & Wave L2 ($\times 10^{-3}$ ) $\downarrow$ & Amplitude L2 $\downarrow$ & Phase L2 $\downarrow$ & PESQ $\uparrow$ & MRSTFT  $\downarrow$ \\\midrule
\textit{DSP}~\cite{leng2022binauralgrad}  & 1.543 & 0.097 & 1.596 & 1.610 & 2.750  \\
\textit{WaveNet}~\cite{leng2022binauralgrad} & 0.179 & 0.037 & 0.968 & 2.305 & 1.915 \\
\textit{NFS}~\cite{Lee2022NeuralFS} & 0.172 & 0.035 & 0.999 & 1.656 & 1.241 \\
\midrule
\textit{WarpNet$^*$}~\cite{Richard2021NeuralSO} & 0.164 & 0.040 & 0.805 & 1.935 & 2.051 \\
\textit{WarpNet$^*$ + DopplerBAS} & \textbf{0.154} & \textbf{0.036} & \textbf{0.780} & \textbf{2.161} & \textbf{2.039} \\ 
\midrule
\textit{BinauralGrad$^*$}~\cite{leng2022binauralgrad} & 0.133 & 0.031 & 0.889 & 2.659 & 1.207 \\
\textit{BinauralGrad$^*$ + DopplerBAS} & \textbf{0.131} & \textbf{0.030} & \textbf{0.869} & \textbf{2.699} & \textbf{1.202} \\
\bottomrule
\end{tabular}
\end{center}
\caption{The comparison regarding binaural audio synthesis quality. For\textit{ WarpNet$^*$} and \textit{BinauralGrad$^*$}, 
we reproduced the results using their official codes (Section~\ref{subsec:setup}).
}
 \vspace{-3mm}
\label{tab:main_res}
\end{table*}

\subsection{DopplerBAS}
\label{sec:method:dopplerbas}
We do not directly apply Eq.~\eqref{eq:doppler} in the frequency domain of audio, because some previous works~\cite{Lee2022NeuralFS} show that modeling the binaural audio in the frequency domain 
degrades the accuracy although it could benefit the generalization ability.
Different from modeling the Doppler effect in the frequency domain, we calculate the velocity of interest and use it as a condition to guide the neural network to synthesize binaural audio consistent with the moving event. In the receiver-centric Cartesian coordinates, we define $\vec{p}_s$ and $\vec{p}_e$ as the 3D position of the moving sound source $s$ and one ear of the receiver $e$ respectively (e.g., the right ear, as shown in Figure~\ref{fig:velocity_decompose}). The position vector $\vec{p} = (p_x, p_y, p_z) $ of $s$ relative to $e$ is:
\begin{align*}
    \vec{p} = (p_x, p_y, p_z) = \vec{p}_s - \vec{p}_e .
\end{align*}
Then $s$'s velocity~\footnote{This velocity is the same in all the Cartesian coordinate systems relatively stationary to the receiver.} can be calculated as:
\begin{align*}
    \vec{v} = (v_x, v_y, v_z) = (\frac{\mathrm{d} p_x }{\mathrm{d} t}, \frac{\mathrm{d} p_y }{\mathrm{d} t}, \frac{\mathrm{d} p_z }{\mathrm{d} t}) .
\end{align*} 
Next, we build the spherical coordinate system using the ear as the origin, and decompose $\vec{v}$ into the radial relative velocity $\vec{v}_r$ by:
\begin{align}
\label{eq:velocity}
    \vec{v}_r = \frac{\vec{p} \cdot \vec{v}}{\| \vec{p} \|} \cdot \hat{\mathbf{r}},
\end{align}
where $\hat{\mathbf{r}} \in \mathcal{R}^1$ is the radial unit vector. 

Finally, we add $\vec{v}_r$ as the additional condition to the network: The original conditions in monaural-to-binaural speech synthesis are $C_o \in \mathcal{R}^7 = (x, y, z, qx, qy, qz, qw)$, of which the first 3 represent the positions and the last 4 represent the head orientations. We define the new condition $C \in \mathcal{R}^9 = (x, y, z, qx, qy, qz, qw, v_{r-left}, v_{r-right})$, where $v_{r-left}$ and $v_{r-right}$ represent the radial velocity of source relative to the left and right ear respectively, which are derived from Eq.~\eqref{eq:velocity}. We then apply $C$ to WarpNet and BinauralGrad backbones, as follows.

\subsubsection{WarpNet} 
\label{sec:method:warpnet}
WarpNet consists of two blocks: 
1) The Neural Time Warping block to learn a warp from the source position to the listener's left ear and right ear while respecting physical properties~\cite{Richard2021NeuralSO}. This block is composed of a geometric warp and a parameterized neural warp. 
2) The Temporal ConvNet block to model subtle effects such as room reverberations and output the final binaural audio. This block is composed of a stack of hyper-convolution layers. 
We replace the original $C_o$ with $C$ for the input of parameterized neural warp and for the condition of hyper-convolution layers.

\subsubsection{BinauralGrad} BinauralGrad consists of two stages:
1) The ``Common Stage'' generates the average of the binaural audio. The conditions for this stage include the monaural audio, the average of the binaural audio produced by the geometric warp in WarpNet~\cite{Richard2021NeuralSO}, and $C_o$. 
2) The ``Specific Stage'' generates the final binaural audio. The conditions for this stage include the binaural audio produced by the geometric warp, the output of the ``Common Stage'', and $C_o$. 
BinauralGrad adopts diffusion model for both stages, which is based on non-causal WaveNet blocks~\cite{vanwavenet} with a conditioner block composed of a series of 1D-convolutional layers. We replace $C_o$ with $C$ as the input of the conditioner block for both stages.

\section{Experiments}
In this section, we first introduce the commonly used binaural dataset, and then introduce the training details for WarpNet-based and BinauralGrad-based models. After that, we describe the evaluation metrics that we use to evaluate baselines and our methods. Finally, we provide the main results with analytical experiments on BAS.

\subsection{Setup}
\label{subsec:setup}
\paragraph{Dataset}
We evaluate our methods on the standard binaural dataset released by \citet{Richard2021NeuralSO}. It contains 2 hours of paired monaural and binaural audio at 48kHz from eight different speakers. Speakers were asked to walk around a listener equipped with binaural microphones. An OptiTrack system track the positions and orientations of the speaker and listener at 120Hz, which are aligned with the audio. We follow the original train-validation-test splits as \citet{Richard2021NeuralSO} and \citet{leng2022binauralgrad} for a fair comparison.


\paragraph{Training Details}
We apply DopplerBAS on two open-source BAS systems WarpNet and BinauralGrad. We train 1) WarpNet and WarNet+DopplerBAS on 2 NVIDIA V100 GPUs with batch size 32 for 300K steps, and 2) BinauralGrad and BinauralGrad+DopplerBAS on 8 NVIDIA A100 GPUs with batch size 48 for 300K steps~\footnote{Following the recommended training steps in their official repository.}. 



\paragraph{Evaluation Metrics}
Following the previous works~\cite{leng2022binauralgrad,Lee2022NeuralFS}, we adopt 5 metrics to evaluate baselines and our methods: 1) \textbf{Wave L2}: the mean squared error between waveforms; 2) \textbf{Amplitude L2}: the mean squared errors between the synthesized speech and the ground truth in amplitude; 3) \textbf{Phase L2}: the mean squared errors between the synthesized speech and the ground truth in phase; 4) \textbf{PESQ}: the perceptual evaluation of speech quality; 5) \textbf{MRSTFT}: the multi-resolution spectral loss.

\subsection{Main Results and Analysis}

\paragraph{Main Results}
We compare the following systems: 1) \textit{DSP}, which utilizes the room impulse response~\cite{Lin2006BayesianRA} to model the room reverberance and the head-related transfer functions~\cite{Cheng2001IntroductionTH} to model the acoustic
influence of the human head; 2) \textit{WaveNet}~\cite{Richard2021NeuralSO,leng2022binauralgrad}, which utilizes the WaveNet~\cite{vanwavenet} model to generate binaural speech; 3) \textit{NFS}, which proposes to model the binaural audio in the Fourier space; 4) \textit{WarpNet}~\cite{Richard2021NeuralSO}, which proposes a combination of geometry warp and neural warp to produce coarse binaural audio from the monaural audio and a stack of hyper-convolution layers to refine coarse binaural audio; 5) \textit{WarpNet + DopplerBAS}, which applies \textit{DopplerBAS} to \textit{WarpNet}; 6) \textit{BinauralGrad}~\cite{leng2022binauralgrad}, which proposes to use diffusion model to improve the audio naturalness; 7) \textit{BinauralGrad + DopplerBAS}, which applies \textit{DopplerBAS} to \textit{BinauralGrad}.

The results are shown in Table~\ref{tab:main_res}. ``\textit{+ DopplerBAS}'' could improve both \textit{WarpNet} and \textit{BinauralGrad} in all the metrics, especially in the Phase L2 metric. \textit{WarpNet + DopplerBAS} performs best in the Phase L2 metric and reaches a new state of the art \textbf{0.780}. \textit{BinauralGrad + DopplerBAS} obtains the best Wave L2, Amplitude L2, PESQ and MRSTFT score among all the systems. These results show the effectiveness of \textit{DopplerBAS}.

\paragraph{Analysis}
We conduct analytical experiments for the following four velocity conditions. ``\textit{Spherical $\vec{v}$} '': the velocity conditions introduced in Section~\ref{sec:method:dopplerbas} are calculated in the spherical coordinate system; ``\textit{Cartesian $\vec{v}$} '': the velocity conditions are calculated in the Cartesian coordinate system; ``\textit{Zeros}'': the provided conditions are two sequences of zeros; ``\textit{Time series}'': the provided conditions are two sequences of time. The results are shown in Table~\ref{tab:ablation}, where we place WarpNet in the first row as the reference. We discover that: 
1) Radial relative velocity is the practical velocity component, which obeys the theory of the Doppler effect (row 2 vs. row 1); 
2) The velocity condition is beneficial to binaural audio synthesis, even for the absolute velocity in the Cartesian coordinates (row 3 vs. row 1); 
3) Just increasing the channel number of the condition $C_o$ (Section~\ref{sec:method:dopplerbas}) by increasing the parameters in neural networks without providing meaningful information could not change the results (row 4 vs. row 1); 
4) The neural networks do not explicitly learn the derivative of position to time (row 5 vs. row 1). These points verify the rationality of our proposed method.

\begin{table}[t]
    \centering
    \footnotesize
    \begin{tabular}{l|l|ccc}\toprule
        No. & \textbf{Model} &
        W. L2 & Amp. L2 & Phase L2 \\
        \midrule
        1 & \textit{WarpNet} & 0.164 & 0.040 & 0.805 \\
        \midrule
        2 & \textit{+Spherical $\vec{v}^\dagger$}  & \textbf{0.154} & \textbf{0.036} & \textbf{0.780} \\
        3 & \textit{+Cartesian $\vec{v}$}         & 0.164 & 0.038 & \underline{0.790}  \\
        4 & \textit{+Zeros}         & 0.159 & 0.038 & 0.806   \\
        5 & \textit{+Time series}   & 0.163 & 0.039 & 0.822   \\
        \bottomrule
    \end{tabular}
    \caption{Analysis Experiments. ``W. L2'' means Wave L2 $\cdot10^3$; ``Amp. L2'' means Amplitude L2; $^\dagger$ means our method: \textit{DopplerBAS}. Best scores over the corresponding baseline are marked in bold.}
    \label{tab:ablation}
\end{table}

\section{Conclusion}
In this work, we proposed DopplerBAS to address the Doppler effect of the moving sound source in binaural audio synthesis, which is not explicitly considered in previous neural BAS methods. We calculate the radial relative velocity of the moving source in the spherical coordinate system as the additional conditions for BAS. Experimental results show that DopplerBAS scales well to different types of backbones and reaches a new SOTA. Analyses further verify rationality of DopplerBAS.


\section*{Limitations}
The major limitation is that we test our method only on a binaural speech dataset, in which there is a person moving slowly while speaking. Because this person moves slowly, the Doppler effect is not so obvious. We will try to find or collect a sound dataset of a source moving at high speed, such as a running man, flying objects, or vehicles, and further, analyze the experimental phenomena at different speeds of the moving source.

\section*{Ethics Statement}
The immersive experience brought by space audio may make people indulge in the virtual world.

\section*{Acknowledgements}
This work was supported in part by the National Key R\&D Program of China under Grant No.2022ZD0162000,National Natural Science Foundation of China under Grant No.62222211, Grant No.61836002 and Grant No.62072397. This work was also supported by Speech Lab of DAMO Academy, Alibaba Group.

\bibliography{custom} 
\bibliographystyle{acl_natbib}



\end{document}